\documentstyle[11pt,newpasp,twoside,epsf]{article}

\begin{document}

\title{A CS J$= 5 \rightarrow 4$ Mapping Survey Towards High Mass 
       Star Forming Cores Associated with H$_2$O Masers}
\author{Y. L. Shirley, N. J. Evans II, K. E. Mueller, C. Knez, 
        \& D. T. Jaffe}
\affil{Department of Astronomy, The University of Texas at Austin,
       Austin, Texas 78712--1083}

\begin{abstract}
In this survey we have systematically mapped 63 cloud cores 
in the CS J$= 5 \rightarrow 4$ line towards a sample of high mass
star forming cores with water masers (Plume et al. 1992, Plume et al. 1997)
using the CSO.  From the CS spectra and maps we determine 
cloud core sizes, virial masses, a mass spectrum, and a 
size-linewidth relationship.  
\end{abstract}

\section{Introduction}

	Sixty-three high mass star forming cores 
selected from the study of Plume et al. (1997; $T_R^* > 1$K) 
were mapped in CS J$= 5 \rightarrow 4$ at the CSO 
between September 1996 and July 1999.
Nearly all of the cores observed lie in the first and second
galactic quadrants.  The cores were mapped using the
on-the-fly technique with a square grid in RA-DEC coordinates oversampled
at 10$\arcsec$ resolution.  The maps were extended until the CS emission
was negligible ($\pm 50\arcsec$ for the average map).

\section{Discussion}

	The size of each core was determined by deconvolving the 
telescope main beam FWHM from the observed FWHM of the core. 
Fifty-seven cores had clearly defined FWHM with an average size
of $0.38 \pm 0.28$ pc.  Six cores contained multiple peaks within
the core FWHM and were not included in the average.  
Thirty-three cores had deconvolved sizes that are larger than the CSO 
beamsize indicating that the majority of cores were resolved.
The distribution of core sizes (Figure 1a) is peaked about the
mean; however, we are biased against small cores sizes due to the
resolution of the CSO beam and the large distance of high
mass star forming regions ($<D> = 5.5 \pm 3.7$ kpc).

	A size-linewidth relationship can be determined from 
Gaussian fits to the linewidth of the convolved map of the source
($<\Delta v> = 5.78 \pm 2.74$ km/s).
The size-linewidth relationship shows a weak correlation 
($r_{corr} = 0.54$; Figure 1b) with a least squares fit of
$\Delta v \sim R^{0.76 \pm 0.07}$ and 
robust estimation fit of $\Delta v \sim R^{0.34}$.  All of
the linewidths are significantly higher than the size-linewidth 
relationship determined by Caselli \& Myers (1995) for massive cores
in Orion indicating that the average high mass star forming core linewidth
associated with a water maser is very turbulent.

	The mean virial mass of a core was calculated using the
CS linewidth and core FWHM size to be $4550 \pm 1380$ M$_{\sun}$ 
which translates into an average surface density of $1.68$ g cm$^{-2}$. 
The corresponding mass spectrum is well fit by a power law (Figure 1c), 
$dN/dM \sim M^{-1.8}$ to $M^{-2.0}$ using least squares and
robust estimation.  Since we are biased against detecting
small core masses ($M < 1000$ M$_{\sun}$), the power laws fit
to the mass spectrum is restricted to masses larger than 10$^3$
solar masses.  The virial mass was derived assuming a constant density
envelope and that the CS line was optically thin.
Preliminary results from dust modeling of 350 $\micron$ emission ($n \sim r^{-p}$, 
$<p> = 1.73 \pm 0.35$ for 28 cores; Mueller et al. these proceedings)
and C$^{34}$S observations ($\Delta v($CS$)/\Delta v($C$^{34}$S$) = 1.3 \pm 0.3$ 
for 10 cores) indicate that the average virial mass and
surface density are reduced by a factor of 0.43 when these effects are
included.

\begin{figure}
\plotfiddle{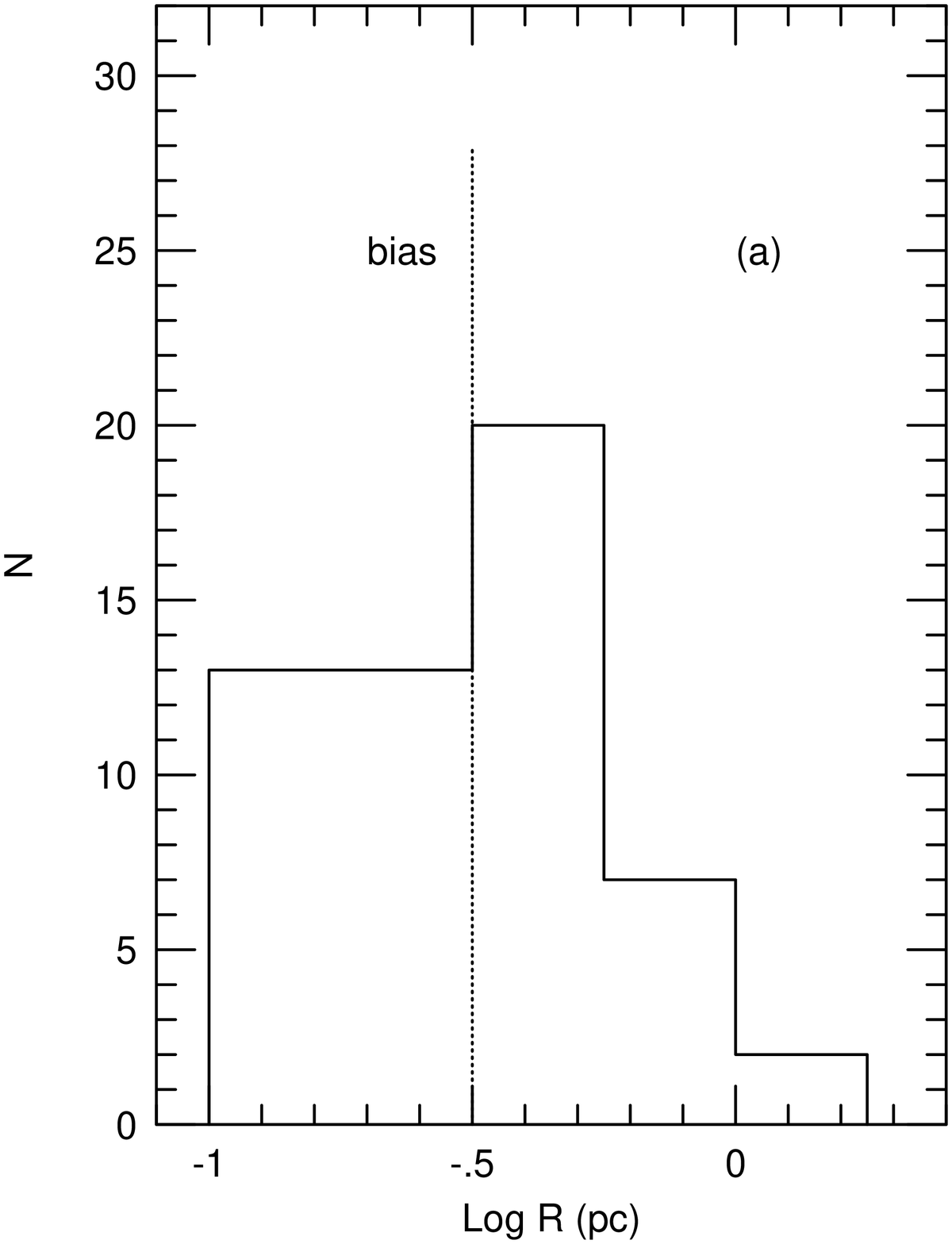}{2.0in}{90}{25}{25}{-60}{-40}
\plotfiddle{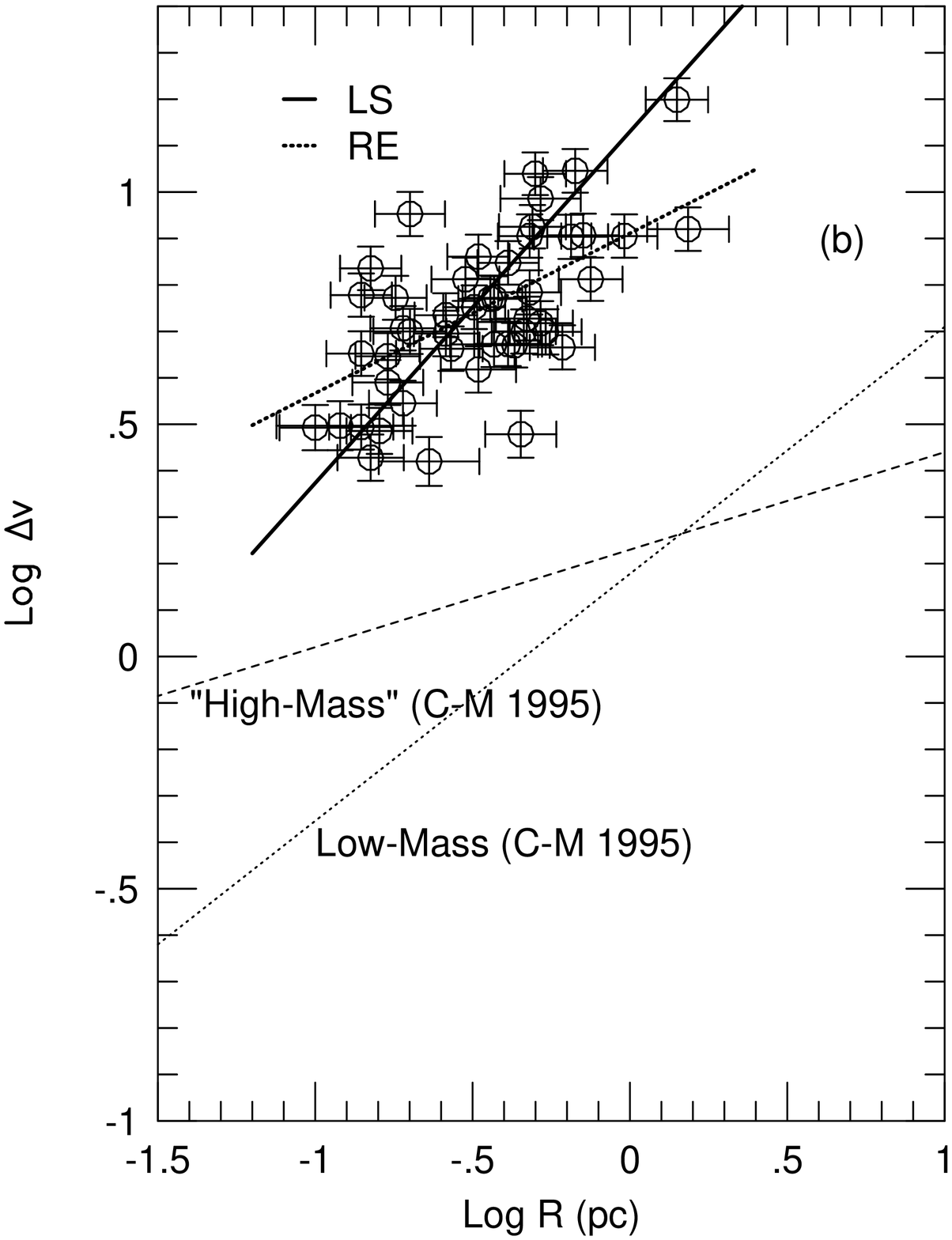}{0.0in}{0}{25}{25}{-77}{-100}
\plotfiddle{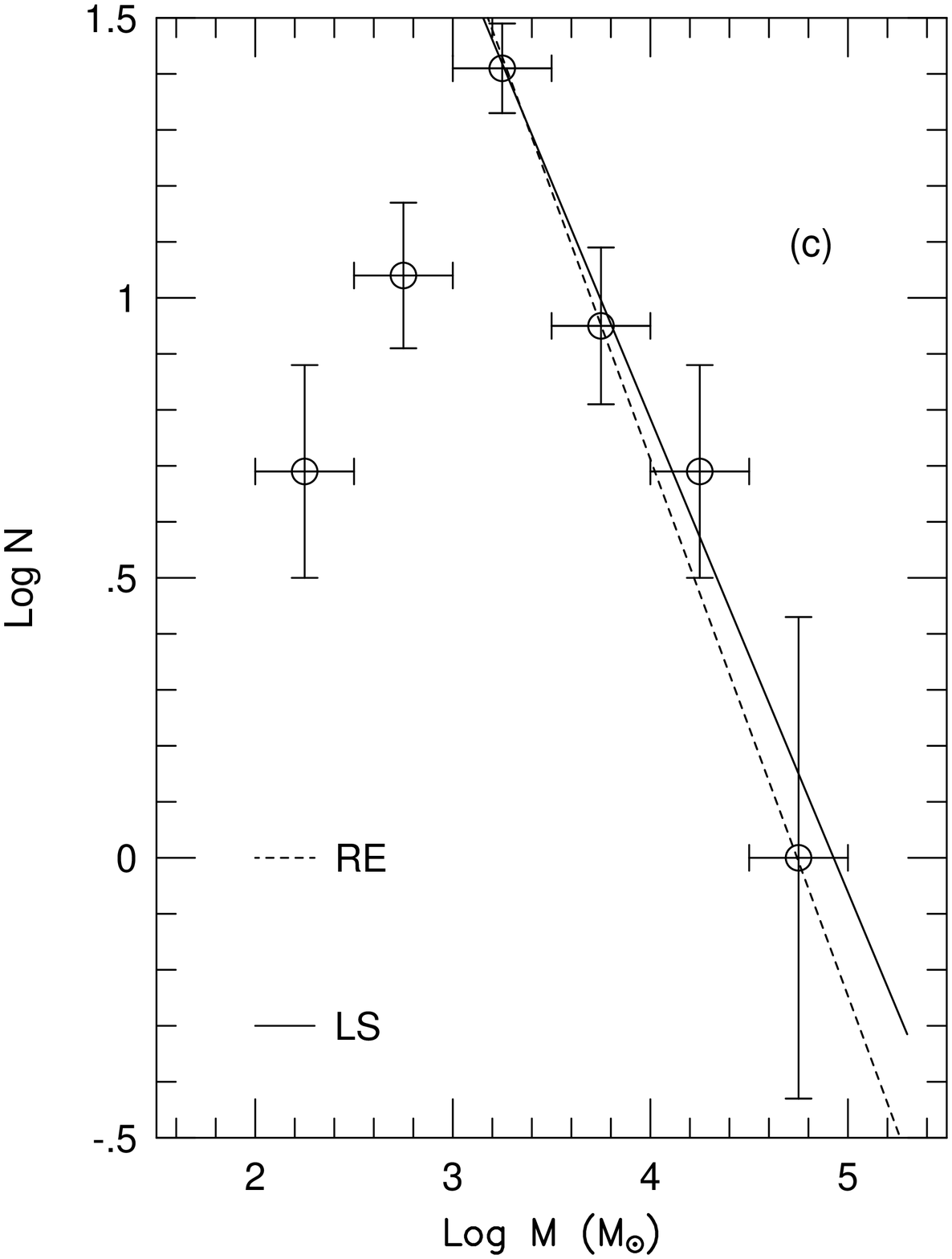}{0.0in}{0}{25}{25}{60}{10}
\caption{The size histogram (a), size-linewidth relationship (b), and mass
spectrum (c) for 57 cores. LS refers to least squares fit and RE 
refers to robust estimation.}
\end{figure}

	This CS mapping survey provides a more unbiased look at the statistical
properties of high mass star forming cores associated with water masers.
Combined with studies of dust continuum emission (Mueller et al. these 
proceedings) and Monte Carlo modeling of multiple CS
transitions (Knez et. al. these proceedings), a more
definitive picture of the density and temperature structure of
the envelopes of high mass star forming cores is emerging.

\end{document}